\documentclass[twocolumn]{webofc}
\usepackage[varg]{txfonts}   % Web of Conferences font

\begin{document}
\message{The text height is \the\textheight and this the \the\footskip}
\title{ANNI -- A pulsed cold neutron beam facility for particle physics at the ESS}
%
% subtitle is optional

\author{\firstname{Torsten} \lastname{Soldner}\inst{1}\fnsep\thanks{\email{soldner@ill.fr}} \and
        \firstname{Hartmut} \lastname{Abele}\inst{2} \and
        \firstname{Gertrud} \lastname{Konrad}\inst{3,2} \and
	\firstname{Bastian} \lastname{M{\"a}rkisch}\inst{4} \and
	\firstname{Florian M.} \lastname{Piegsa}\inst{5} \and
	\firstname{Ulrich}  \lastname{Schmidt}\inst{6} \and
	\firstname{Camille} \lastname{Theroine}\inst{1,4,7} \and
	\firstname{Pablo}   \lastname{Torres S{\'a}nchez}\inst{1,8}
        % etc.
}

\institute{Institut Laue-Langevin, CS 20156, 38042 Grenoble Cedex 9, France
\and
           Atominstitut, TU Wien, Stadionallee 2, 1020 Wien, Austria
\and
           Stefan Meyer Institute for Subatomic Physics, Austrian Academy of Sciences, 
	   Boltzmanngasse 3, 1090 Wien, Austria
\and
           Physik-Department ENE, Technische Universit{\"a}t M{\"u}nchen, James-Franck-Str. 1, 85748 Garching, Germany
\and
           Laboratory for High Energy Physics and Albert Einstein Center for Fundamental Physics,
	   University of Bern, Sidlerstrasse 5, 3012~Bern, Switzerland
\and
           Physikalisches Institut, Universit{\"a}t Heidelberg, Im Neuenheimer Feld 226, 69120 Heidelberg, Germany
\and
           European Spallation Source, Box 176, 22100 Lund, Sweden
\and
           Fac Ciencias, Universidad de Granada, Av. Fuente Nueva S/N, 18071 Granada, Spain
          }

\abstract{%
   Pulsed beams have tremendous advantages for precision experiments with cold neutrons.
   In order to minimise and measure systematic effects, they are used at continuous sources
   in spite of the related substantial decrease in intensity. At the European Spallation
   Source ESS these experiments will profit from the pulse structure of the source and its
   50 times higher peak brightness compared to the most intense reactor facilities, making
   novel concepts feasible. Therefore, the cold
   neutron beam facility for particle physics ANNI was proposed as part of the ESS instrument
   suite. The proposed design has been re-optimised to take into account the present ESS cold
   moderator layout. We present design considerations, the optimised instrument parameters
   and performance, and expected gain factors for several reference experiments.
}
\maketitle
\section{Introduction}
\label{Sec:Introduction}

Neutron particle physics is part of the precision frontier in particle physics.
New particles and interactions are not observed directly but detected by their
contributions to observables at low energy. In this way, experiments with neutrons
can probe new physics at the multi-TeV scale, complementary to and competitive with
searches at colliders \cite{Abele2008,Dubbers2011,Cirigliano2013,GonzalesAlonso2019,Chupp2018}.
Precision is key in this endeavour. Experimental progress requires high statistics
and a flawless control of systematic effects. Important experimental conditions can be
controlled by using pulsed beams:
\begin{description}
  \item[Neutron velocity (energy, wavelength):] The arrival time of a neutron at a given
    distance from the creation of the pulse indicates its velocity.
    This can be exploited to separate effects with different dependencies on the
    neutron velocity. Examples are the beam polarisation and
    transmission by a spin filter cell which can be used for precise polarimetry \cite{Pentilla2005}
    or the induced magnetic field $\propto {\bf v}\times{\bf E}$ in the rest frame of
    a neutron moving in an electric field which is relevant for measurements of the
    neutron electric dipole moment (EDM) \cite{Piegsa2013}. In a novel neutron interferometric
    concept for a precision measurement of the electric charge of the neutron
    the timing information is used to select neutrons of specific velocities \cite{Piegsa2018}.
    Furthermore, parameters of neutron-velocity dependent optical elements can be
    varied to fit the neutrons arriving at a certain moment
    (\textbf{time-dependent neutron optics}). Examples are the amplitude of the oscillatory fields
    in $\pi/2$ spin-flip coils enabling a $\pi/2$ flip for all neutrons \cite{Piegsa2013}
    or the opening of a disk chopper effectively working as beam monochromator.
  \item[Localisation of neutron pulse in time:] At a pulsed beam the signal arrives
    within a short time span which increases the signal-to-background ratio compared to
    a continuous beam of the same time-averaged intensity, assuming similar levels
    of environmental background. Furthermore, signal and background are measured in
    time intervals that are very close to each other, allowing to subtract 
    slowly-changing environmental background \cite{Maerkisch2014}. Beam-related
    background with a time  constant large compared to the pulse duration (e.g.\ activation
    of windows or beta-active
    isotopes in a target, background from unwanted Penning traps in retardation spectrometers)
    is diluted compared to the prompt signal which increases the signal-to-background ratio
    for beam-related background.
  \item[Localisation of neutron pulse in space:] In contrast to a continuous beam, a neutron pulse
    can be observed without any contact to baffles or the beam stop thus preventing beam-re\-la\-ted
    background, or fully inside a region with well-defined spectrometer response hence
    avoiding e.g.\ edge effects \cite{Maerkisch2014}.
\end{description}
These conditions reduce or completely suppress many systematic effects. Therefore
pulsed beams are used at continuous facilities in spite of the related substantial
reduction in counting statistics. At a pulsed facility, these methods can be applied
at full counting statistics. The ESS cold moderator will provide a 50 times higher
peak brightness and a two times higher time-averaged brightness than the cold sources
of the leading continuous facility (the high flux reactor of the Institut Laue-Langevin
ILL) \cite{Andersen2018}. In spite of the substantially smaller viewed area of the ESS
cold moderator, pulsed experiments at the ESS should gain more than
one order of magnitude in counting rate compared to the ILL. This led us to propose
ANNI \cite{Theroine2015}, a pulsed cold neutron beam facility at the ESS optimised
for precision measurements of neutron beta decay, hadronic weak interaction in
calculable systems \cite{Gardner2017}, and electromagnetic properties of the neutron.

\section{Design Considerations}
\label{sec:DesignConsiderations}

The design of ANNI is driven by the goal to provide the maximum usable neutron
flux as well as all advantages of pulsed beams. In order to provide unique velocity
information, frame overlap must be absent for the accepted part of the spectrum.
The wavelength band of 2~\AA{} to 8~\AA{} is chosen as reference as it includes
90\%{} of the cold capture flux from the moderator. These requirements call for
a short instrument which can be sensitive to background from the spallation
source itself. Therefore, the next prompt pulse shall not fall inside a signal
period at ANNI, i.e.\ arrive after the slowest selected neutrons have left the
instrument. With the ESS repetition rate of 14~Hz, the signal area of an
experiment at ANNI should not surpass 34~m distance from the moderator. Note
however that the usable
signal area depends on the chosen wavelength band and that some experiments
may be insensitive to background from the prompt pulse. Further
arguments support the choice of a short instrument: The neutron pulse
dilutes with distance from the moderator which makes its localisation 
in time less effective. The statistically optimal length of
the beam line for a PERC-like spectrometer \cite{Dubbers2008} is about 20~m
(position of the last chopper) \cite{Klauser2014}. At about 22~m from the source,
the instantaneous bandwidth (difference of longest and shortest wavelength present
at a given distance and time) for the full source pulse falls below 0.5~\AA{}
which enables a wavelength resolution of below 10\%{} for cold neutrons.
On the other hand, ANNI shall be versatile and be able to host long experiments
such as the proposed BeamEDM \cite{Piegsa2013} (length of the interaction area
about 50~m).

For pulse localisation the last chopper should be as close as possible
to the experiment. In order to fit long spectrometers such as
PERC \cite{Dubbers2008} (8~m decay volume) in front of the 34~m argued
above, the last chopper is placed at 26~m from the moderator. The ANNI guide
ends at 22~m, leaving 4~m space for beam preparation.
Because of the high-energy particles created in the source,
the ESS recommends that guide design excludes direct sight to the moderator
at least twice (where subdividing walls in multi-channel benders are not
taken into account), in order to prevent showers created in the first guide contact
of these particles from reaching the experiment directly. The ANNI guide design
follows this recommendation. This requires a strongly curved guide. The ESS
moderator is flat which calls for a guide with smaller vertical than
horizontal dimensions. Therefore the smallest guide curvature is achieved by
curving in the vertical plane. In order to obtain a horizontal beam,
an S-shaped geometry is used in the vertical plane.

Because of the small moderator, there is a delicate tradeoff between flux density
and integral flux at the guide exit which was optimised using a suite of
reference experiments. The resulting guide has $O(10~\text{cm})$ height.
In order to achieve a critical wavelength of about 2~\AA{} with economical
supermirrors and a homogeneous beam profile in spite of the strong curvature,
the curved sections are realised by multi-channel benders. The two
benders have the same characteristic parameters $\gamma^*$ and $\lambda^*$
(defined in e.g.\ \cite{Dubbers1994})
but were chosen to be different: The second is relatively compact, with
a total length of 2.5~m, in order to be able to replace it by a polarising
bender. This allows for the highest polarised intensity if the
requirements for the absolute polarisation are moderate. The instrument
design also includes options for ultra-high polarisation.

ANNI's guide and chopper design was optimised for a selection of
reference experiments covering the various beam configurations used at
existing cold neutron beam facilities. These reference experiments were
aSPECT \cite{Zimmer2000,Gluck2005},
NPDGamma \cite{Gericke2011,Blyth2018}, PERC \cite{Dubbers2008} and
Perkeo\,III \cite{Maerkisch2009}. Details on the selection of these
experiments and the configurations used in the simulations are given in
appendix~\ref{app:ReferenceExperiments}. The BeamEDM experiment
\cite{Piegsa2013} was taken into account for defining the spatial
requirements of ANNI. The new projects BRAND \cite{Bodek2018}
and HIBEAM \cite{Young2018} have expressed interest to measure at
ANNI.

\section{Instrument Layout}

\subsection{Simulations for Guide Optimisation}

The final simulations for the ANNI instrument proposal were performed in
2015 for the former 3~cm flat ``Pancake'' moderator design \cite{Zanini2015}.
Technically, the simulations were performed with McStas \cite{Willendrup2011}
version 2.1 using the source component \verb|ESS_moderator| (with
\verb|sourcedef="2014"| and the solid angle bug-fix from February 2015)
and the guide component \verb|Guide_gravity|.
Benders were composed of 16 straight guide elements with small gaps
corresponding to a frequent practical realisation of long benders. Gravity
was switched off in the simulations and geometrical guide imperfections
were ignored. The reflectivities of guide supermirrors were described by
realistic $m$-dependent parameters (where $m$ is the supermirror factor,
example parameters for $m=3$: \verb|R0=0.99|, \verb|Qc=0.0219|,
\verb|W=0.0015|, \verb|alpha=2.74| in McStas notation).

The final ``Butterfly'' moderator BF1 \cite{Andersen2018} has the same
height of 3~cm, but contains a thermal section. This reduces the
effective width of the cold part to about 8~cm for beam port E5,
the beam port provisionally allocated to ANNI \cite{Andersen2016},
compared to about 17~cm of the ``Pancake'' moderator. Simulations
for this moderator were performed with McStas version 2.4 and the
source component \verb|ESS_butterfly|; the beam port was
specified by \verb|sector="E"| and \verb|beamline=5|.
Each instrument can orient the guide to the thermal or the cold
section of the moderator or opt for bispectral extraction \cite{Andersen2018}.
For ANNI the cold section can be selected by
starting the guide at $x=4.3$~cm and aligning it parallel to the beam
port axis or by starting the guide at $x=0$ and aligning it at
$-1.26^\circ$ to the beam port axis (all in McStas coordinates). In
practice, an intermediate solution may best fit the guide into the
beam port; according to McStas simulations this has negligible influence
on the performance.
For this paper, only the horizontal guide dimensions were re-optimised
whereas the vertical geometry was kept unchanged compared to the proposal
\cite{Theroine2015}. The results presented here base on the simulations
for the Butterfly moderator.

\subsection{Guide Layout}

\begin{figure*}
  \centering
  \includegraphics[width=\textwidth,clip]{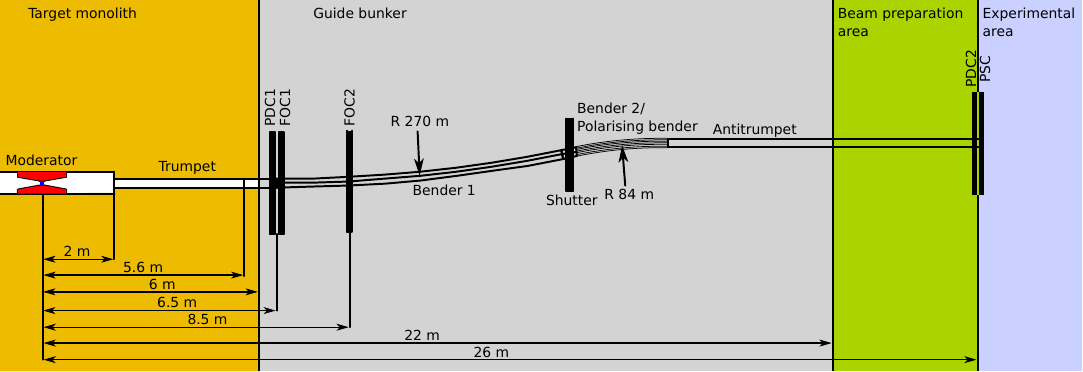}
  \caption{Side view of the ANNI beam line (schematic). The vertical scale
    is stretched by a factor of 4 for better readability. See text and
    tables~\ref{tab:guide} and \ref{tab:choppers} for details.}
  \label{fig:SideView}
\end{figure*}
A vertical cut through the ANNI guide is shown in figure~\ref{fig:SideView}.
Table~\ref{tab:guide} lists the key parameters of all guide components.

\begin{table*}
  \caption{Key parameters of the guide components. Dist: Distance from the moderator to
    the start of the element. Rad: Radius of curvature. Negative means bend downwards. Len: Length
    of element. X-section: $\text{width}\times\text{height}$. Trumpet and antitrumpet
    change the cross-section linearly from start to end (indicated by arrow).
    NOC: Number of channels in vertical direction.
    Channels are separated by 1~mm thick plates. $m$ value: supermirror coating for all
    sides or separately for Left, Right, Top, and Bottom seen from the moderator. 
    $\gamma^*$ and $\lambda^*$: characteristic parameters of curved guides.}
  \label{tab:guide}
  \centering
  \begin{tabular}{lcccccccc}
    \hline
    Element & Dist   & Rad& Len& X-section                      & NOC & $m$ value & $\gamma^*$ & $\lambda^*$ \\
            & [m]    & [m]     & [m]    & [cm$^2$]                           &     &           & [mrad]     & [\AA] \\\hline
    Trumpet &    2.0 & $\infty$& 3.6    & $9\times6\rightarrow13\times6$ & 1   &LR: 3.5, TB: 3.0&       &       \\
    Straight 1 & 5.6 & $\infty$& 0.9    & $13\times6$                      & 1   &3.0             &       &       \\
    Bender 1 &   6.5 & 270     & 8.0    & $13\times6$                      & 2   &LRT: 3.0, B: 3.5& 14.7  & 2.44  \\
    Straight 2 &14.5 & $\infty$& 0.4    & $13\times6$                      & 2   &3.0             &       &       \\
    Bender 2 or Po-
             &   6.5 & -84     & 2.5    & $13\times6$                      & 6   &LRB: 3.0, T: 3.5& 14.7  & 2.44  \\
    larizing bender\\
    Antitrumpet& 17.4& $\infty$& 4.6    & $13\times6\rightarrow11\times7$& 1   &LR: 3.5, TB: 3.0&       &       \\\hline
  \end{tabular}
\end{table*}

\begin{figure*}
  \centering
  \includegraphics[width=\textwidth,clip]{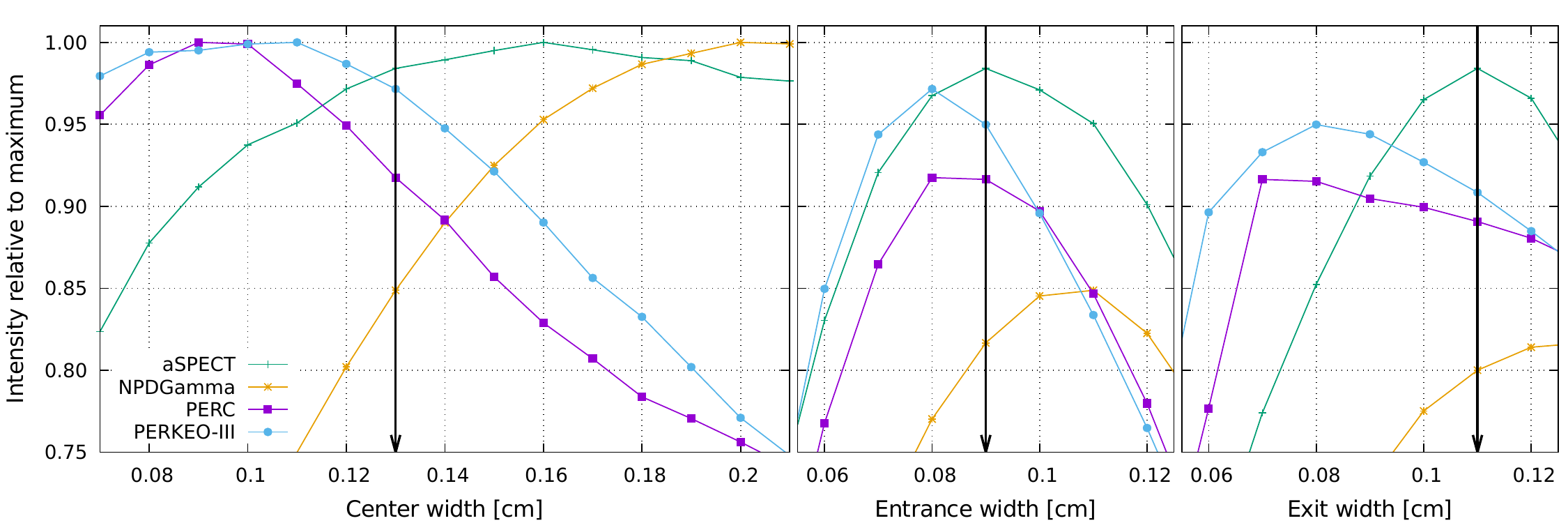}
  \caption{
    Optimisation of central, exit and entrance width of the guide system.
    All other guide settings were fixed to the values given in
    table~\ref{tab:guide}. Intensities are
    plotted relative to the maximum for the respective reference experiment
    (see appendix~\ref{app:ReferenceExperiments})
    in the full parameter space. In the left plot, the maximum intensities
    for a given central width and free entrance and exit widths are plotted.
    The arrow indicates the selected central width (13~cm) which is then
    fixed for the other two plots.
    In the middle plot, the maximum intensities for a given entrance width
    and free exit width are plotted; the arrow indicates the selected
    entrance width (9~cm) which is then fixed for the right plot. The right
    plot shows the intensity as function of the exit width (other widths
    fixed) and is used to select the exit width of 11~cm. Although the
    positions of the maxima for the different reference experiments are
    quite different, for the selected compromise the intensities for all
    reference experiments are above  80\%{} of the respective maximum.
    This best compromise can be different if one wants to optimise for
    a certain class of experiments. Note that neither the flux density nor
    integral flux at the guide exit are good criteria for this optimisation
    (the flux density can be maximised by minimising the exit width, resulting
    in a large divergence, and the integral flux by a wider guide, resulting
    in reduced flux density) and are therefore not plotted.}
  \label{fig:OptimizationWidth}
\end{figure*}

As illustration of the optimisation procedure, figure~\ref{fig:OptimizationWidth}
shows the selection of the central, entrance and exit width of the guide
from the simulations with the Butterfly moderator.
The optimisation in the vertical plane is more complex since the curvature
has to be taken into account. According to the simulations it is preferable
to expand the guide from 6~cm to 7~cm height only at the end, after the
curved section, which is counter-intuitive in terms of ballistic transport
but allows for larger radii of curvature.

\subsection{Floor plan}

\begin{figure*}
  \centering
  \includegraphics[width=\textwidth,clip]{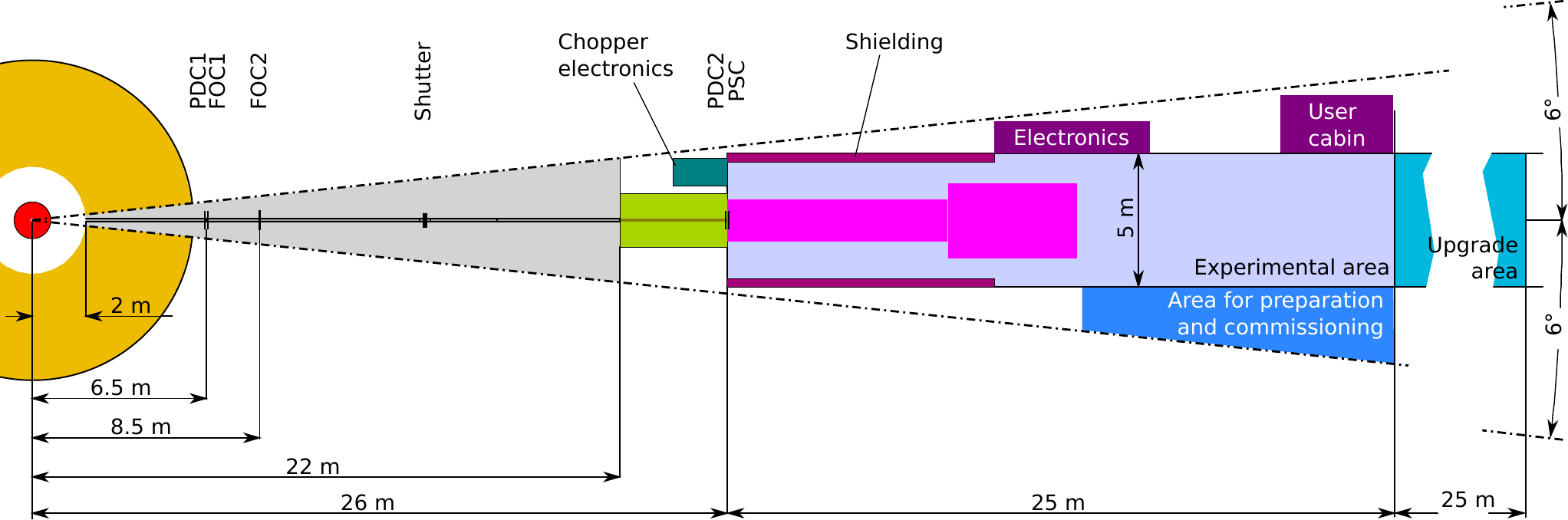}
  \caption{Proposed floor plan of the ANNI facility (schematic).
    Standard colors are as in figure~\ref{fig:SideView}. To give a scale,
    the magenta area indicates a PERC-like instrument with secondary
    spectrometer in measuring position. The  upgrade area is needed
    for long experiments such as BeamEDM.}
  \label{fig:TopView}
\end{figure*}
The proposed floor plan is shown in figure~\ref{fig:TopView}.

The guide ends at 22~m from the moderator and is followed by a 4~m long
beam preparation area where neutron optical elements can be installed
to shape the beam for a specific experiments. This includes
additional polarisers, flippers, focussing optics, straight
guides to transport the full flux and collimation systems. The beam
preparation area shall be protected by a casemate but accessible
if the shutter is closed.

The experimental area of ANNI starts at 26~m from the center of the moderator.
We propose a width of 5~m for the experimental area, which corresponds
to $\pm5.5^\circ$ at 26~m. The angle between beam ports being 6$^\circ$, the
left and right neighbor beam ports should not be allocated (or their instruments
sufficiently downstream). The experimental area should be 25~m long. We propose
to reserve the full angular sector of $\pm6^\circ$ between the neighbor beam
ports for ANNI as this would allow to prepare and commission the next experiment
close to the measuring position, reducing time for changeovers. In order to
realise the BeamEDM experiment \cite{Piegsa2013}, the experimental area
should be extendable to 50~m length which is possible at the provisionally
allocated beam port E5 \cite{Andersen2016}. 

\subsection{Chopper system}
\label{subsec:ChopperSystem}

The chopper system consists of two frame overlap choppers (FOC1, FOC2),
two pulse-defining choppers (PDC1, PDC2) and one pulse-suppressing chopper PSC.
The PDCs are composed of pairs of co-rotating disks in order to adapt the opening
angle to the experiment. Running the PDCs at multiples of the ESS repetition rate
14~Hz (e.g.\ at 70~Hz) and using two counter-rotating pairs instead of one single 
pair per PDC greatly
increases the rate in the experiments since chopper transition times (time
intervals of only partly opened beam during the passage of a chopper slit) are
strongly reduced and the chopper openings can be larger for a given
pulse localisation \cite{TorresSanchez2018}.
The PSC is needed to suppress the higher order pulses of the PDCs.

Pulse multiplication can be applied at ANNI by replacing the PDC2 by a PMC, which
is a multi slit chopper with slits at different angular distances from each other
and of different sizes. The additional pulses appear at different neutron wavelength
which enables certain systematic checks for each source pulse. Since this PMC
has to rotate at the ESS repetition rate, two counter-rotating disks are necessary
in order to obtain acceptable chopper transition times. The layout of the PMC has
to be adapted to the requirements of the specific experiment. Studies for Perkeo\,III
indicate that increasing the rotation frequency of the
standard PDCs (with adapted slit sizes) brings a higher gain in statistics than
pulse multiplication \cite{TorresSanchez2018}, however, the benefit of pulse
multiplication strongly depends on the experiment.

The parameters of the choppers are summarised in table~\ref{tab:choppers}.
\begin{table*}
  \caption{Key parameters of the choppers. Dist: Distance from the moderator,
    Rad: radius of the chopper disk (indicative). Freq: Rotation frequency range
    (used frequencies are multiples of 14~Hz). Angle: opening angle of the
    chopper slit. X-section: beam cross-section at the position of the chopper.
    Choppers are placed left or right of the beam to minimise the transition times.
    (E) indicates chopper disks that shall be compatible with a close-by
    particle physics experiment (non-depolarising, compatible with high vacuum,
    coated with $^6$Li for low gamma background).}
  \label{tab:choppers}
  \centering
  \begin{tabular}{lcccccl}\hline
    Chopper & Dist & Rad & Freq & Angle & X-section & Comment \\
            & [m]  & [cm]   & [Hz]      & [$^\circ$] & [cm$^2$] &      \\\hline
    PDC1  & 6.5 & 40 & 14--70 & Variable & $13\times6$ & 2 counter-rotating pairs of co-rotating disks\\
    FOC1  & 6.5 & 40 & 14     & 64       & $13\times6$ & Single disk \\
    FOC2  & 8.5 & 40 & 14     & 79       & $13\times6$ & Single disk \\
    PDC2 (E) & 26  & 40 & 14--70 & Variable & $11\times7$ & 2 counter-rotating pairs of co-rotating disks\\
    PSC (E)  & 26  & 40 & 14     & 60       & $11\times7$ & Single disk, for PDC2 frequencies > 14 Hz\\
    PMC (E)  & 26  & 40 & 14     & Multi slit & $11\times7$ & 2 counter-rotating disks, replaces PDC2 \\\hline
  \end{tabular}
\end{table*}
The following chopper modes are available:
\begin{description}
  \item[Full intensity:] All choppers in rest in open position. Frame
    overlap occurs.
  \item[Maximum intensity with wavelength information:] Only FOC1
    and FOC2 are running and reject neutrons outside the wavelength band of
    2~\AA{} to 8~\AA{} (the wavelength band can be shifted by changing the
    phase between the choppers). Frame overlap is suppressed up to 130~\AA{}.
  \item[Localisation in time:] Frame overlap is suppressed by the FOCs.
    PDC2 is running with large opening adjusted to the highest intensity
    of the beam. Depending on the wavelength bandwidth, PDC2 is run at a
    high frequency, in combination with the PSC.
  \item[Monochromatic:] All choppers are phased to wavelength $\lambda_0$.
    The PDCs run at high frequency, in combination with the PSC. The opening
    angles of the PDCs are tuned to let all neutrons of wavelength $\lambda_0$
    pass.
  \item[Localisation in space:] As Monochromatic mode, but the opening
    angles and phases of the PDCs are tuned to the dimensions of a spectrometer
    such as PERC or Perkeo\,III (see \cite{Klauser2014,TorresSanchez2018}
    and appendix~\ref{app:ReferenceExperiments}).
\end{description}

\subsection{Polarising Options}
\label{subsec:PolarisingOptions}

Neutron beam polarisation is necessary for many particle physics experiments
where the requirements depend on the experiment: Searches for
non-zero values of small asymmetries require highest intensity and can
trade in a larger uncertainty of neutron beam polarisation, whereas
accurate absolute measurements of asymmetries start to need $10^{-4}$
accuracy on the polarisation value \cite{Dubbers2008,Maerkisch2018}. Therefore
ANNI provides different options for neutron polarisation:
\begin{description}
  \item[Moderate polarisation at highest intensity:] Bender 2 is replaced
    by a polarising supermirror bender of identical dimensions. The only
    additional losses of beam intensity by the polariser are due to the polarisation
    process itself and potentially due to lower reflectivities of polarising
    supermirrors compared to standard supermirrors.
  \item[Highest polarisation:] The polarising supermirror bender
    (see previous point) is combined with a second polarising supermirror
    bender in the beam definition area. The two polarising benders implement
    the X-SM geometry \cite{Kreuz2005}. The neutron spin is rotated
    adiabatically between the two polarisers. For this geometry, an average
    beam polarisation of well above 99.9\% can be expected since the
    second polariser can be compact (e.g.\ solid state polariser \cite{Petoukhov2016})
    and be installed in a high magnetising field, suppressing depolarisation
    by the supermirror \cite{Klauser2016}. In general the beam will be
    deflected horizontally by the second polariser which necessitates a
    lateral translation of the PDC2.
  \item[Polarisation with analytical wavelength dependence:]
    For pulsed beams, the analytical wavelength dependence of the beam
    polarisation after a $^3$He spin filter can be exploited directly for
    precision neutron polarimetry \cite{Pentilla2005}.
    The guide is used in its non-polarising configuration with Bender 2.
    A $^3$He spin filter cell is installed in the beam definition area,
    either a MEOP cell which can already stand the neutron intensity or a SEOP cell
    if the performance losses in high flux \cite{Sharma2008} can be compensated.
    An adiabatic fast passage spin flipper for $^3$He allows inverting the
    spin direction in-situ which provides possibilities for systematic checks.
  \item[Spin flipping:] An adiabatic fast passage spin flipper for neutrons
    or other flipper types can be installed in the beam definition area.
\end{description}
For polarisation analysis, opaque $^3$He spin filter cells in a magnetic
environment with intrinsic $^3$He spin flipping shall be available,
enabling $10^{-4}$ accuracy \cite{Soldner2011,Klauser2016}. The preferred method
for polarising these cells is a central MEOP facility at the ESS, similar
to the Tyrex facility at the ILL \cite{Andersen2005,Petoukhov2006}.

\section{Expected Performance}

\begin{figure*}
  \centering
  \includegraphics[width=0.49\textwidth,clip]{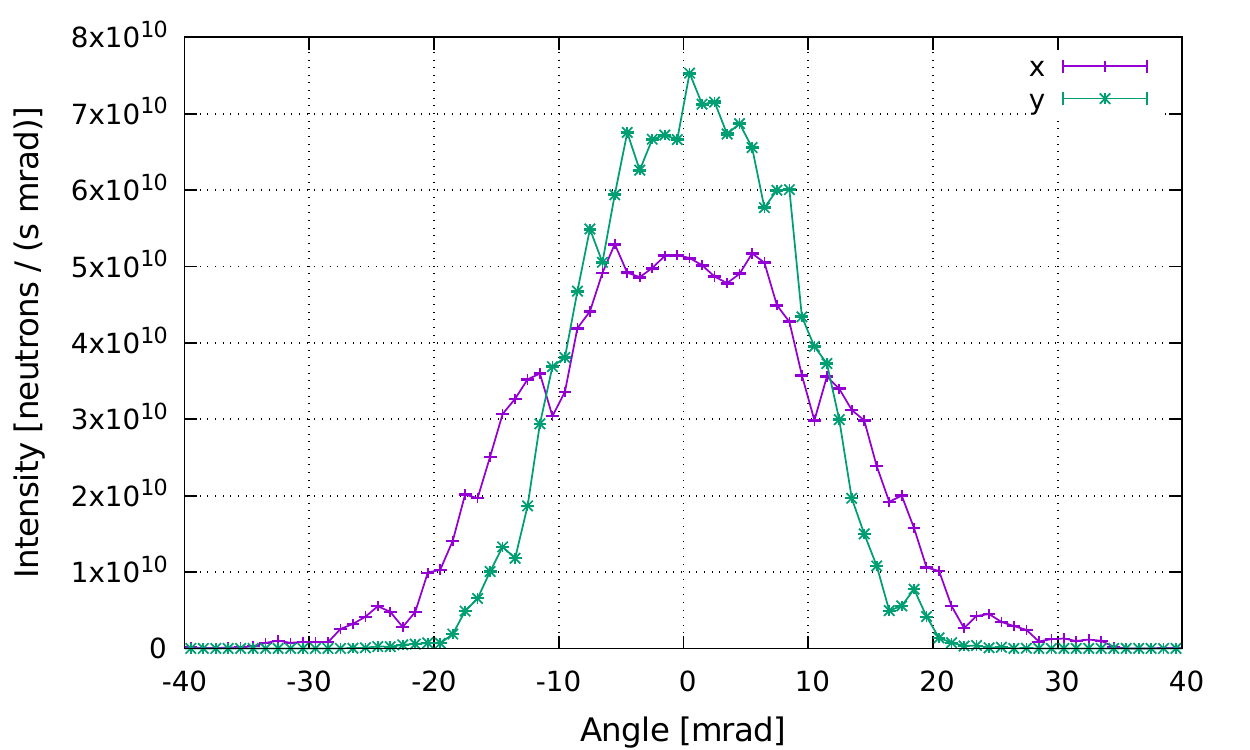}\hfill
  \includegraphics[width=0.49\textwidth,clip]{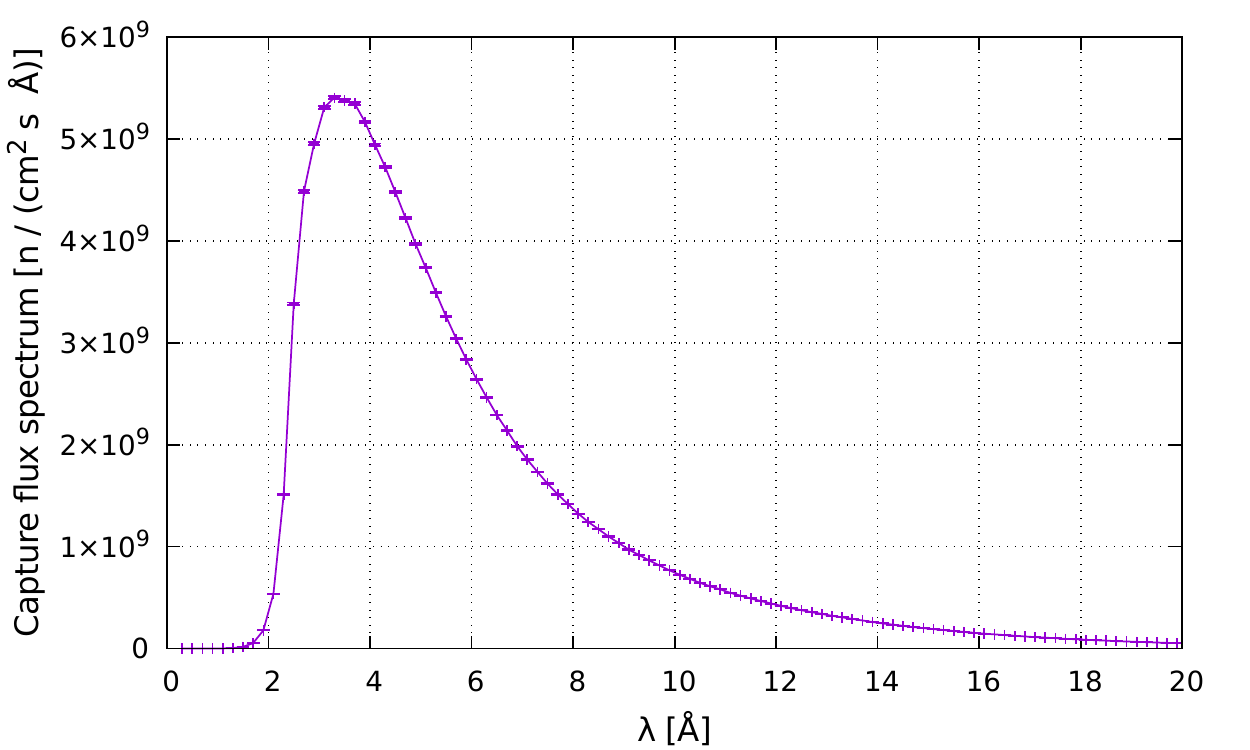}
  \caption{Left: Simulated divergence distributions in horizontal ($x$) and vertical ($y$)
    direction at the guide exit for neutrons between 2~\AA{} and 8~\AA (weighted with their
    capture cross-section and integrated over the other direction and the guide exit area).
    Right: Simulated capture flux density spectrum averaged over the guide exit.}
  \label{fig:DivergencesSpectrum}
\end{figure*}

The simulated divergence distributions in horizontal and vertical direction
and capture flux density spectrum at the guide exit of ANNI are shown in
figure~\ref{fig:DivergencesSpectrum}.
The simulated capture flux density averaged over the guide exit of $11\times 7$~cm$^2$
and time is $2.0\times 10^{10}$~n/(cm$^2$s) for the wavelength band 2~\AA{} to 8~\AA{}
(with FOCs) and $2.5\times 10^{10}$~n/(cm$^2$s) for the full spectrum, comparable
to the capture flux density at the guide exit of the instrument PF1B
\cite{Abele2006} at the ILL ($2.2\times 10^{10}$~n/(cm$^2$s)).
The simulated particle flux density at 8.9~\AA{}, the wavelength relevant for single-phonon UCN
production in He UCN sources \cite{Golub1977}, is $2.1\times 10^8$~n/(cm$^2$s\AA{}).
At ANNI, polarised fluxes for the option of moderate polarisation are only a
factor 2 lower (see section \ref{subsec:PolarisingOptions}), whereas geometrical and
transmission losses of the polarising bender (typically another factor of 2) have
to be taken into account at other facilities.

The expected gain factors for the reference experiments are given in
table~\ref{tab:GainFactors}.

\begin{table}
  \caption{Simulated gain in counting rate for the reference experiments
    at ANNI (ESS at 5~MW) relative to the respective presently used facility.
    Gains marked by $^{\uparrow}$ refer to polarised beams.
    Details on the determination of the gain factors are given
    in appendix~\ref{app:ReferenceExperiments}.}
  \label{tab:GainFactors}
  \centering
  \begin{tabular}{lcl}\hline
    Experiment & Gain & Comment \\\hline
    aSPECT     & 1.3& Full spectrum \\
               & 2.8& Localisation in time to 1/3 of\\
	       &    & ESS period\\
    NPDGamma   & 15$^{\uparrow}$ & Wavelength information\\
    PERC       & 15$^{\uparrow}$ & Localisation in space\\
    Perkeo\,III & 17$^{\uparrow}$ & Localisation in space\\\hline
  \end{tabular}
\end{table}

\section{Conclusions and Outlook}

The proposed neutron beam facility ANNI will provide unprecedented pulsed
intensity for particle physics experiments and its time-averaged flux will
still at least equal existing reactor facilities. Its design allows
to fully exploit the ESS pulse structure and to use the optimum
polarisation option for each experiment. Particle physics experiments
using pulsed beams will gain one order of magnitude in
event rate compared to other facilities. ANNI will enable a new level
of accuracy in measurements of correlation coefficients in neutron
beta decay, novel methods to determine electromagnetic properties of
the neutron and systematic access to the tiny effects of
hadronic weak interaction in calculable systems.

The design proposed in this paper can be further optimised and
should therefore be considered as preliminary. In particular, particle
physics experiments have more demanding background requirements than
instruments for neutron scattering. Therefore the ANNI design should
be validated and optimised
for background suppression from the spallation source. As example, the
long bending sections may be replaced by compact solid state benders,
which may result in more localised sources for secondary particles
that are easier to shield. This would also allow for a compact solid
state polariser \cite{Petoukhov2016} as integrated polariser option
with better performance. Furthermore, the design proposed here is chosen
as best compromise for the considered suite of reference experiments
but scientific priorities could also call for full optimisation towards
one specific experiment.

\textbf{Acknowledgements}
We thank Dr.~Grammer and Dr.~Fo\-min for sharing the MCNPX
model of NPDGamma at the SNS and Mr.~Hol\-le\-ring for
providing McStas parametrisations of the supermirrors
of the PERC internal guide.

\begin{appendix}
\section{Reference experiments}
\label{app:ReferenceExperiments}

The suite of reference experiments was selected in order to cover
most of the particle physics experiments at cold neutron beams. A
short description of the selected experiments is given below;
the most relevant parameters for the simulations are summarised
in table~\ref{tab:ReferenceExperiments}.
\begin{table*}
  \caption{Parameters of the reference experiments used in the simulations.
    The extension guide bridges the distance between the exit of the ANNI
    guide and the start of the collimation system, in order to place the experiment
    itself at the start of the experimental area, and has the same
    cross-section and $m$ value as the ANNI guide. For important apertures
    the distance to the end of the extension guide and the cross-section
    are given. Criterion indicates the quantity that was used for
    optimisation; it is measured after the last aperture. The PERC
    setup is too different and is described in the text.}
  \label{tab:ReferenceExperiments}
  \centering
  \begin{tabular}{lllll}\hline
    Experiment           & aSPECT & NPDGamma & PERC                & Perkeo\,III  \\\hline
    Wavelength range     & 2.0-8.0~\AA & 2.0-8.0~\AA & 4.5-5.5~\AA & 4.5-5.5~\AA \\
    Extension guide      & 1.5~m    &  3.1~m & see text & 0.55~m\\
    First aperture       & as guide exit & as guide exit & see text & $6\times6$~cm$^2$ \\
    Defining apertures   & 2.484~m: $4.5\times 7.0$~cm$^2$ & 0.92~m: $16\times 16$~cm$^2$ &see text & 3.23~m: $6\times6$~cm$^2$\\
                         & 2.894~m: $4.5\times 7.0$~cm$^2$ & 1.72~m: $18\times22$~cm$^2\cap\varnothing 22$~cm \\
    Criterion            & Capture flux &                  Particle flux & see text & Capture flux\\\hline
  \end{tabular}
\end{table*}

\begin{description}
  \item[aSPECT] \cite{Zimmer2000,Gluck2005} is the prototype for a
    neutron decay experiment with short decay volume, low 
    sensitivity to gamma background and consequently a large
    angular acceptance for neutrons. The experiment uses an
    unpolarised continuous beam. Since Penning traps may cause
    background with time
    constants large compared to the ESS pulse length \cite{Maisonobe2014},
    aSPECT may profit from localisation of the neutron pulse in time for
    an improved signal-to-background ratio. From the collimation system
    used in the aSPECT experiment (see \cite{Borg2010}), only the last
    aperture in front of the spectrometer and all apertures inside the
    cryostat were used in the simulations since this is preferable for
    statistics and systematics (smaller edge effect because of the more
    homogeneous beam profile). The gain in table~\ref{tab:GainFactors}
    corresponds to the ratio of the simulated capture intensities in the
    decay volume for aSPECT at ANNI (all choppers at rest, full neutron
    spectrum) and at PF1B. Pulse localisation in time was realised by
    the PDC2 chopper directly upstreams of the last aperture in front
    of the spectrometer. At PF1B, only this chopper was used; at ANNI
    additionally the FOCs suppressed frame overlap. The PDC2 was run at
    14~Hz and its opening was chosen to limit the arrival of neutrons in 
    the decay volume to 1/3 of the chopper period (at PF1B, $10^{-4}$ of
    the intensity were allowed in the other 2/3 since a single chopper
    does not cut the tail of the spectrum). This time localisation reduces
    the time-averaged counting rate at ANNI to about 62\% and at at PF1B to
    about 28\% of the respective full rate.
  \item[NPDGamma] \cite{Gericke2011,Blyth2018} is the prototype
    for target experiments with large target size, large divergence acceptance
    and beam polarisation optimised for intensity.
    Frame overlap suppression is used. The parameters of the collimation
    system were extracted from the MCNPX model \cite{Grammer2017}.
    Since 43\%{} of the neutrons incident on the target are captured by
    hydrogen \cite{Blyth2018}, the particle flux is used as optimisation
    criterion. The gain in table~\ref{tab:GainFactors} corresponds to the ratio
    of the integral particle flux at the guide exit without frame overlap for
    ANNI (FOCs set to 2.5-8.5~\AA{} in order to remain below the 14.7~meV
    threshold for spinflip scattering, see \cite{Grammer2015}) and
    FnPB \cite{Fomin2015} (3.1-6.6~\AA{}, flux from \cite{Blyth2018} scaled to
    1.4~MW nominal power of the Spallation Neutron Source SNS). For the
    FnPB, the geometrical and transmission loss by the subsequent polarising bender
    was taken as 50\% and taken into account; this loss is included in the simulations
    for ANNI with polarising bender 2 implemented in the guide (the factor $1/2$
    for spin selection is the same in both cases).
    The flux at the guide (respective polariser) exit can be used due to the
    high angular acceptance of NPDGamma; the flux at the target position was not available.
    The larger wavelength band at ANNI in spite of the longer guide
    is enabled by the lower repetition rate of the ESS (14~Hz) compared to
    the SNS (60~Hz).
  \item[PERC] \cite{Dubbers2008} is the prototype for experiments where
    the fiducial volume is (almost) directly coupled to the
    primary neutron guide and where the divergence acceptance corresponds to
    that of a neutron guide. The fiducial volume of PERC consists of an 8~m long 
    internal neutron guide where charged neutron decay products are extracted by
    a strong longitudinal magnetic field. The experiment will profit from pulse localisation
    in space in order to avoid regions of ill-defined spectrometer
    response. The beam configuration and the parameters of the internal
    guide were extracted from a McStas model \cite{Hollering2017}.
    For the ANNI simulations, PERC is placed such that the chopper position
    corresponds to that of PDC2. The internal guide must be non-depolarising
    at the $10^{-4}$ level wherefore CuTi supermirrors are used, with $m=1.95$.
    In order to adapt the divergence delivered by the ANNI guide, the beam
    preparation area is bridged by an $m=1.8$ guide of $7.0\times 6.0$~cm$^2$ and 4 m length,
    followed by an 80~cm $m=1.95$ CuTi guide after a 20~cm gap for the PDC2.
    The internal guide starts after a gap of 50~cm (needed for a
    backscattering detector for charged particles). It has 6~cm height and
    consists of through-going top and bottom plates and 4 sections of 1.98~m
    long side plates with 2~cm pumping gap and widths of 7.0~cm, 7.2~cm,
    7.4~cm, and 7.6~cm, respectively. The intensity was determined by
    averaging the capture flux over squares of $5.0\times5.0$~cm$^2$
    between the four sections (the smaller area accounts for the fact
    that decay particles close to the guide surfaces are lost by
    gyration). The guide optimisation was done for a limited wavelength
    band as needed for pulse localisation in space.
  \item[Perkeo\,III] \cite{Maerkisch2009} is the prototype for experiments
    with a long decay volume and limited angular acceptance (about 10~mrad)
    for neutrons.
    The experiment uses pulse localisation in space in order to single
    out decay events in a region with well-defined spectrometer
    response and free from beam-related background. The collimation
    system was taken from \cite{Mest2011} and placed such that the chopper
    position corresponds to that of PDC2. A limited wavelength band
    was used for the guide optimisation in order to simulate pulse
    localisation in space.
\end{description}

The gains for PERC and Perkeo\,III in table~\ref{tab:GainFactors}
are defined as the ratio of the time-averaged decay rates at ANNI and at
PF1B \cite{Abele2006} for the neutron pulse fully contained in the
homogeneous part of the decay volume (pulse localisation in space).
Intensity losses for polarisation were taken into account as described for
NPDGamma above.  At ANNI, pulse localisation was achieved by running
both PDCs at 70~Hz; the other parameters of the PDCs were optimised for
maximum count rate. FOCs and PSC were used to suppress unwanted pulses.
At PF1B, pulse localisation was achieved by combining a Dornier velocity
selector \cite{Friedrich1989} with a chopper. For Perkeo\,III the 
single disk chopper used in the experiment was simulated (see \cite{Mest2011});
the resulting decay rate agreed with the experimental one. For PERC a pair of
counter-rotating disk choppers was used; selector rotation frequency,
chopper slit size and chopper frequency were optimised for maximum time-averaged
decay rate with the condition of at least 2~ms for background measurement between
subsequent pulses, without neutrons between chopper and beam stop (at ANNI this
condition is fulfilled due to the low ESS repetition rate). The statistically
optimal neutron pulse length at the end of the PERC decay volume was found to be
5~m for both ANNI and PF1B. The performance of the new dedicated PERC beam line
MEPHISTO \cite{Mephisto2015} at the FRM-II is expected to be similar to PF1B.
The expected gains for unpolarised measurements would be a factor 2 lower
and are comparable to the simplified estimates in \cite{Klauser2014}.
Further details on the simulations can be found in \cite{TorresSanchez2018}
(where slightly smaller horizontal dimensions of the ANNI guide were used).

\end{appendix}

%For figure with side-caption legend use syntax of figure
%\begin{figure}
%% Use the relevant command for your figure-insertion program
%% to insert the figure file.
%\centering
%\sidecaption
%\includegraphics[width=5cm,clip]{ESSTopView4.png}
%\label{fig-3}       % Give a unique label
%\end{figure}

%
% BibTeX or Biber users please use (the style is already called in the class, ensure that the "woc.bst" style is in your local directory)
% \bibliography{name or your bibliography database}

\vspace{-2cm}
\end{document}